\documentclass[aps,twocolumn,showpacs]{revtex4}
\usepackage{color}
\usepackage{graphicx}
\usepackage{graphics}
\usepackage{epsfig}
\usepackage{subfigure}
\usepackage{latexsym}
\usepackage{amsfonts}
\usepackage[english]{babel}
\usepackage[latin1]{inputenc}
\usepackage[all]{xy}
\usepackage{amsmath}
\usepackage{amssymb}
\newcommand{\be}{\begin{equation}}
\newcommand{\ee}{\end{equation}}
\newcommand{\ben}{\begin{eqnarray}}
\newcommand{\een}{\end{eqnarray}}
\newcommand{\bes}{\begin{subequations}}
\newcommand{\ees}{\end{subequations}}

\newcommand{\ov}{\overline}
\newcommand{\bb}{\bibitem}

\def \P{{\cal P}_{01}}

\def \p{\partial}
\def \pr{\prime}
\def \W{{\cal W}}

\def \bb{\bibitem}
\newcommand{\ba}[1]{\begin{array}{#1}}
\newcommand{\ea}{\end{array}}
\newcommand{\bea}[1]{\begin{equation}\left\{\begin{array}{#1}}
\newcommand{\eea}{\end{array}\right.\end{equation}}

\newcommand{\Ref}[1]{(\ref{#1})}
\begin{document}
\title{Constructing networks of defects with scalar fields}
\author{V.I. Afonso$^{a}$, D. Bazeia$^{a}$, M.A. Gonzalez Le\'on$^{b}$, L. Losano$^{a}$, and J. Mateos Guilarte$^{c}$}
\affiliation{{$^{a}$Departamento de  F\'\i sica, Universidade Federal da Para\'\i ba, Brazil}
\\{$^{b}$Departamento de Matem\'atica Aplicada, Universidad de Salamanca, Spain}
\\{$^{c}$Departamento de F\'\i sica and IUFFyM, Universidad de Salamanca, Spain}}

\begin{abstract}
We propose a new way to build networks of defects. The idea takes advantage of the deformation procedure recently employed to describe defect structures, which we use to construct networks, spread from small rudimentary networks that appear in simple models of scalar fields. 
\end{abstract}
\maketitle


Networks are of great interest in physics in general. In high energy physics, networks appear in diverse contexts, usually in scenarios which require the presence of topological defects, as junctions of domain walls \cite{DW}, cosmic strings \cite{S}, and brane tiling \cite{hanany}. The presence and evolution of domain walls and domain wall networks have been investigated in several ways in {\cite{sugra,net,J}, and the dynamical evolution of domain wall networks in an expanding universe has been recently studied in computer simulation in Ref.~\cite{ammmo}.

In the present Letter we focus attention on kink networks, that is, we deal with models described by scalar fields, which develop spontaneous symmetry breaking of discrete symmetry \cite{DW,J}. We then take advantage of the deformation procedure introduced in \cite{DD}, and extended to other scenarios in \cite{EDD},
to deform a given model, described by a potential containing a rudimentary set of minima, to get to another model, with the potential giving rise to a different set of minima, which may replicate periodically. As a bonus, the deformation procedure also gives the defect structures of the deformed model in terms of the defect solutions of the original model. Thus, in the lattice of minima of the deformed model we can nest a network of defects in a very natural way. 

This is the main idea underlying this paper, in which we use the deformation method to investigate two important possibilities, one described by a single real scalar field, giving rise to linear networks, and the other by a complex scalar field, giving rise to planar networks. We focus mainly on the generation of kink-like networks described by the deformed models, which are generated from simple models, which engender rudimentary networks.

The idea of constructing networks of defects is not new, but the novelty here relies on the use of the deformation procedure as a simple and natural way to generate networks. The mechanism is powerful and suggestive, and fully motivates the present work. To make the reasoning mathematically consistent we consider a model described by the Lagrange density with a single real scalar field $\chi$ in the form
\be
{\cal L}=\frac12 \partial_\mu\chi\partial^\mu\chi-\frac12 W^{\prime2}(\chi)
\ee
The potential $V(\chi)=(1/2)W^{\prime2}(\chi)$ is given in terms of the superpotential $W=W(\chi),$ with the prime standing for the derivative with respect to the argument, e.g. $W^\prime(\chi)=dW/d\chi.$ In this case, the equation of motion for static field $\chi=\chi(x)$ can be reduced to the first-order differential equation ${d\chi}/{dx}=W^\prime(\chi)$. For $W=\pm(\chi-\chi^3/3)$ we get the $\chi^4$ model, which has the set of minima $\{-1,1\}$. In this case, the defect structure represents kink $(\tanh(x))$ or anti-kink $(-\tanh(x))$, with energy minimized to the value $E=4/3.$ For simplicity, we are working with dimensionless fields, space-time coordinates, mass and coupling constants, with mass and coupling constants set to unit. In the one dimensional field space, the orbit is a straight line segment which connects the two minima. Since the kink or anti-kink spans the orbit in the positive or negative sense, they may orient the orbit, leading to orientable networks.

We now use an extension of the deformation procedure considered in the first work in \cite{EDD}. The deformed model is described by
\be
{\cal L}_D=\frac12\partial_\mu\phi\partial^\mu\phi-\frac12{\cal W}^{\prime2}(\phi)
\ee
with the deformed potential $U(\phi)$ given in terms of the new superpotential ${\cal W}(\phi)=W(f(\phi))/f^{\prime}(\phi).$ Here $f(\phi)$ is the deformation function, and we consider $f(\phi)= \sin(\phi),$ with inverse $f_n^{-1}(\phi)=(-1)^n {\rm Arcsin}(\phi)+n\pi,$ with $n=0,\pm1,\pm2,...$ This gives another model, the sine-Gordon model with ${\cal W}^{\prime}(\phi)=\cos(\phi).$ The set of minima is now given by $\{(2n-1)\pi/2,(2n+1)\pi/2\}$. It forms a lattice in the entire field space, and $n=0$ identifies the central sector with minima $\{-\pi/2,\pi/2\}$, $n=1$ the sector $\{\pi/2,3\pi/2\},$ and $n=-1$ the sector $\{-3\pi/2,-\pi/2\}$, etc. The orbit of the original model is now mapped into distinct orbits of the new model, giving rise to a specific network, which appears as a spreading of the original set of two points into the entire field space, the real line in the present case. This is illustrated in Fig.~1.  
 
The above study allows the construction of a regular lattice, in which pairs of adjacent minima are equally spaced and connected by kinks and anti-kinks with the very same profile and the same energy $E_D=2.$ We can change regularity of the lattice changing the deformation function. We take for instance $f_a(\phi)=\cos(\phi^a),$ with $a$ real and positive, $a=1$ leading us to a model similar to the former model. It introduces the potential
\be
V(\phi)=\frac1{2a^2}\phi^{2(1-a)}\sin^2(\phi^a)
\ee
In this case, the set of minima is given by $\bar\phi_n=\pm(n\pi)^{1/a},$ $n=0,1,2,...,$ and the distance between consecutive minima in the lattice increases for $a<1$ and decreases for $a>1,$ as we get away from the central minimum at the origin. This case gives another tiling, for which the distance between minima and the corresponding defect energy vary in a nice way, controlled by the parameter $a.$ The energy for $a=2/3$ in the sector labeled by $n$ is now $E^{2/3}_D=(9/4)(2n+1)\pi,$ which increases linearly with $n.$ See \cite{blmm} for further details.

\begin{figure}[htbp]
\centerline{\epsfig{file=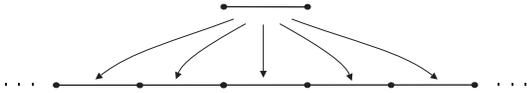,width=7cm}}
\caption{Plot of the minima of the original and deformed potentials, showing how the deformation function projects topological sectors
in the two models.}
\end{figure}

Let us now move to the plane, considering another model, described by a single complex field, $\chi(x,t)=\chi_1(x,t)+i\chi_2(x,t),$ written in terms of the two real fields $\chi_1(x,t)$ and $\chi_2(x,t).$ The specific model which we consider is described by the Lagrange density
\be
{\cal L}=\frac12\partial_\mu\chi\partial^\mu\ov\chi-\frac12 W'(\chi)\,{\ov{W'(\chi)}}
\ee
where the bar stands for complex conjugation. We specify the model choosing $W(\chi)$ as the holomorphic function
\be\label{W3}
W(\chi)=\chi-\frac1{N+1}{\chi^{N+1}}
\ee
This is the Wess-Zumino model. It was investigated before in \cite{V,AT,SP}. The case with $N=3$ is interesting and illustrative: the vacua manifold has the three points $\bar{\chi}_k={\rm exp}(2\pi i(k-1)/3),$ with $k=1,2,3,$ which depict an equilateral triangle in the field plane. And the static solutions satisfy the first-order ordinary differential equation
\be\label{eq:1}
\frac{d\chi}{dx}=e^{i \alpha}\,{\ov{W'(\chi)}}=e^{i\alpha}(1-{\ov\chi}^{3}(x))
\ee
together with the accompanying complex conjugate, where $e^{i\alpha}\in{\mathbb S}^1$ is a phase. We can write $W_\alpha(\chi)=e^{-i\alpha}W(\chi)$ to get $d(W_\alpha-\overline{W}_\alpha)=0.$ This implies that the kink orbits arise when the imaginary part of the superpotential is constant 
\be\label{knkorb}
{\rm Im}\left({e^{-i\alpha}}\left(\chi(x)-{\chi^{4}(x)}/4\right)\right)={\rm const}
\ee
As the kink orbits connect minima of the potential, this constant must also be equal to the value of ${\rm Im}\,W_\alpha$
at those minima, which are the roots of unity. This means that $\sin(2(k-1)\pi/3-\alpha)={\rm constant}.$ 
Of course, this constant value should be the same at the two different minima connected by the orbit. Then we have
${\rm Im}\,W_{\alpha^{(kj)}}(\chi^{(k)})={\rm Im}\,W_{\alpha^{(kj)}}(\chi^{(j)}),$ and so
\be
\alpha^{(kj)}=-{\rm arcsin}\left({\rm cos}\left({\pi}(k+j-2)/3\right)\right),\quad k>j \label{eq:4}
\ee
Note that $\alpha^{(jk)}=\alpha^{(kj)}+\pi$ if $j<k$.

We can also use the first-order equations to obtain 
\be\label{eq:6}
\frac{ds}{dx}=\left|W^\prime(\chi(x))\right|^2=\left|(1-\chi^{3}(x))\right|^2
\ee
where $s$ stands for the ``length" on the kink orbits \Ref{knkorb} -- see Ref.~{\cite{SP}}. The kink profiles are then obtained by inverting these relations between the real part of the superpotential and $s.$ The energy of the static configurations is $E=(3/2)\left|\sin((k-j)\pi/3)\right|.$

We now turn attention to the deformation procedure. We follow the second work in \cite{EDD}. It is interesting to express 
the deformed system in terms of another complex field, $\phi(x,t)=\phi_1(x,t)+i\phi_2(x,t),$ related to the original one by means of a holomorphic function
$f=f(\phi)$ such that
\bes\ben
\chi=f(\phi)=f_1(\phi_1,\phi_2)+if_2(\phi_1,\phi_2)
\\ 
\frac{\p f_1}{\p\phi_1}=\frac{\p f_2}{\p\phi_2};\;\;\;\;\;\frac{\p f_1}{\p\phi_2} =-\frac{\p f_2}{\p\phi_1}
\een\ees
The deformed Lagrange density has the form 
\be
{\cal L}_D=\frac12\p_\mu\phi\p^\mu\ov{\!\phi}- \frac{V(f(\phi),\ov{f(\phi)})}{f^\pr(\phi)\ov{f^\pr(\phi)}} \label{eq:def}
\ee
The deformed model ${\cal L}_D$ can be defined by the new superpotential
\be
\W^\prime(\phi)=\frac{W^\pr(f(\phi))}{\ov{f^\prime(\phi)}}
\ee
In this case, the ``deformed" first-order equations are
\be
\frac{d\phi}{dx}=\,e^{i \alpha}\,\overline{\W^\pr(\phi)};\;\;\;\;\; 
\frac{d\bar{\phi}}{dx}=\,e^{-i\alpha}\,\W^\pr(\phi) \label{eq:12}
\ee
The defect solutions for this system are obtained from the solutions of (\ref{eq:1}) by simply taking the inverse of the deformation function: $\phi^K(x)=f^{-1}(\chi^K(x))$. Thus, we can establish the following relation between the deformed and original equations: if $\chi^K(x)$ is a kink-like solution of the original model, we have that
\be\label{eq:defo1}
{\rm Im}\, W(\chi^K(x))={\rm const};\;\;\;{\rm Re} \, W(\chi^K(x))=s
\ee
and then $\phi^K(x)=f^{-1}(\chi^K(x))$ is kink-like solution of the deformed model, obeying
\be\label{eq:defo2}
{\rm Im}\W(f^{-1}(\chi^K(x)))\!=\!{\rm const};\;\;
{\rm Re}\W(f^{-1}(\chi^K(x)))\!=\!{\sigma}
\ee
where $\sigma$ is defined by
\be
{\sigma}=\int\, |\W^\pr\left(f^{-1}(\chi^K(x))\right)|^2\, dx
\ee

Although the method is general, we now specify the deformed model choosing $f(\phi)=\W(\phi).$ This constrains the function $f(\phi)$ to obey the equation
\be 
f'(\phi) \overline{f'(\phi)}= \sqrt{2V(f(\phi),\overline{f(\phi)})}
\label{eq:sdsu}
\ee
A function $f$ satisfying this condition provides a potential $U(\phi,\overline{\phi})$ for the new model 
which is well defined (finite) at the critical points of $f(\phi),$ e.g. the zeros of $f^\prime(\phi)$. 
As a bonus, the procedure leads to a very simple expression for the deformed superpotential. We change $\chi$ for $f(\phi)$
in the general expression \eqref{W3} to get the potential 
\be
V=\frac12(1-f^N(\phi))(1-\overline{f^N(\phi)})
\ee
As stated in \Ref{eq:sdsu}, we can then choose
\be
f^{\prime\,2}(\phi)=(-1)^N(1-f^N(\phi))
\ee
The solution of this equation solves the general problem. 

We illustrate the general results with $N=3.$ Here we have
\be
f^{\prime\,2}(\phi)=f^3(\phi)-1
\ee 
The solution is the equianharmonic case of the Weierstrass ${\cal P}$ function
\be
{\cal W}(\phi)=f(\phi)=4^{\frac{1}{3}}\,{\cal P}(4^{-\frac{1}{3}}\phi; 0,1)
\ee
The Weierstrass ${\cal P}$ function is defined as the solution of the ODE
\be
({\cal P}^\prime(z))^2=4{\cal P}^3(z)-g_2{\cal P}(z)-g_3 
\ee
The elliptic function which solves the differential equation above is doubly periodic function defined as the series
\be
{\cal P}(z)=\frac{1}{z^2}+\sum_{m,n}\biggl(\frac{1}{(z-A(m,n))^2}-\frac{1}{A(m,n)^2}\biggr)
\ee
where $A(m,n)=2m\omega_1+2n\omega_3,$ with $m,n\in\mathbb{Z}$ and ${m^2+n^2\neq 0}.$ Therefore, the deformation function is, up to a factor, the Weierstrass ${\cal P}$ function with invariants $g_2=0$ and $g_3=1$, and we denote it by ${\cal P}_{01}(z)$.
This function is meromorphic, with an infinite number of poles congruent to the irreducible pole of order two in the fundamental period parallelogram (FPP).

Here we get $\W_\alpha(\phi)=e^{-i \alpha}\,4^{\frac13} \,\P(4^{-\frac13}\,\phi),$ and so the deformed potential can be written as
\be
U(\phi,\overline{\phi})=\frac{1}{2} \P^\pr(4^{-\frac{1}{3}}\phi)\,\overline{\P}\,^\pr(4^{-\frac13}\phi)
\ee
The potential spans the plane replicating the triangular structure as shown in Fig.~2, and in Fig.~3.

The new potential is doubly periodic with an structure inherited from the ``half-periods" of ${\cal P}$. The set of zeros of
the potential in the FPP has three elements $\phi^{(1)}=\omega_1=\omega_2(1/2-i\sqrt{3}/2)$, $\phi^{(2)}=\omega_3=\omega_2(1/2+i\sqrt{3}/2)$,
and $\phi^{(3)}=\omega_2=4^{\frac13}\Gamma^3(1/3)/{4\pi}$. The set of all the zeros of $U$ form a lattice which tile the entire field plane,
as we show in Fig.~3.

\begin{figure}[htbp]
\centerline{\epsfig{file=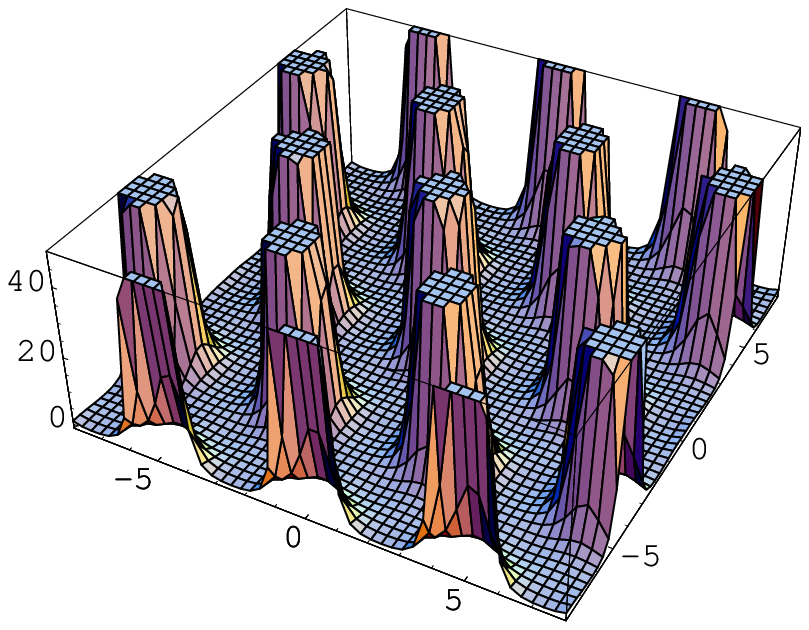,height=3.6cm}}
\centerline{\epsfig{file=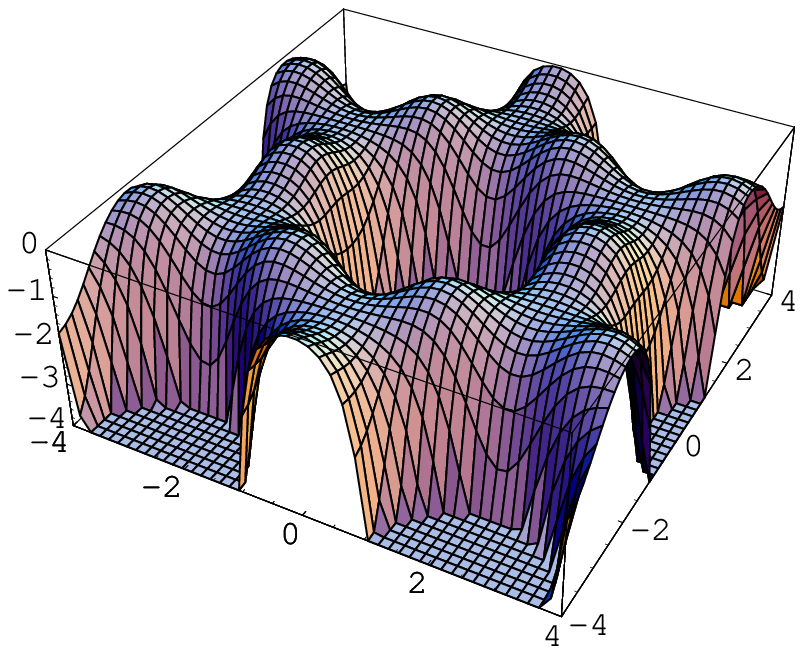,height=3.6cm}}
\caption{The case $N=3,$ showing the potential $U(\phi,{\bar\phi})$ (upper panel) and its mechanical analogue $-U(\phi,{\bar\phi})$ near a pole (lower panel). Note that in the lower panel the zeros are now maxima.}
\end{figure}

The potential obtained from the deformation procedure has the same number of zeros in the FPP as the original model in the whole field space. Besides, one pole of sixth order arises at the origin due to the meromorphic structure of ${\cal P}^\prime_{01}(4^{-\frac13}\phi);$ see Fig.~3. However, this structure is infinitely repeated in the deformed model, according to the two periods $\omega_1$ and $\omega_3$ determining the modular parameter $\tau={\omega_3}/{\omega_1}=-{1}/{2}+i\sqrt{3}/2$ of the Riemann surface of genus 1 associated with this ${\cal P}$-Weierstrass function.

We now compare the ${\cal P}$-kink orbits with the orbits of the original model. If $\chi^K(x)$ is a
solution of (\ref{knkorb}) and (\ref{eq:6}) then $\phi^K(x)=4^\frac13\P^{-1}(4^{-\frac13}\chi^K(x))$ solves
\begin{eqnarray}
{\rm Im}\;e^{-i\alpha}\,4^{\frac13}\,\P(4^{-\frac13}\phi^K(x))&=&{\rm const} \nonumber\\ 
{\rm Re}\;e^{-i\alpha}\,4^{\frac13}\,\P(4^{-\frac13}\phi^K(x))&=&\sigma
\label{eq:d6}
\end{eqnarray}
where
\be
\sigma=\int\,\left|\P^\pr(4^{-{\frac13}}\phi^K(x))\right|^2\, dx
\ee

Since the deformation function is a conformal transformation, angles are preserved, and the same values of $\alpha$ as in the non deformed case give the kink orbits. There are three types, and here we just inform that the nearest neighbor type (12), (23) and (31) minima are connected by orbits which follow specific sequences -- see Ref.~{\cite{abglmNPB}}. We notice that the kink orbits go around, circumventing the singularities which stand at the center of circles depicted by the orbits themselves. Like in the case of a single real scalar field, we can also break the lattice regularity in this case -- see \cite{abglmNPB} for further details on this issue.

\begin{figure}[htbp]
\centerline{\epsfig{file=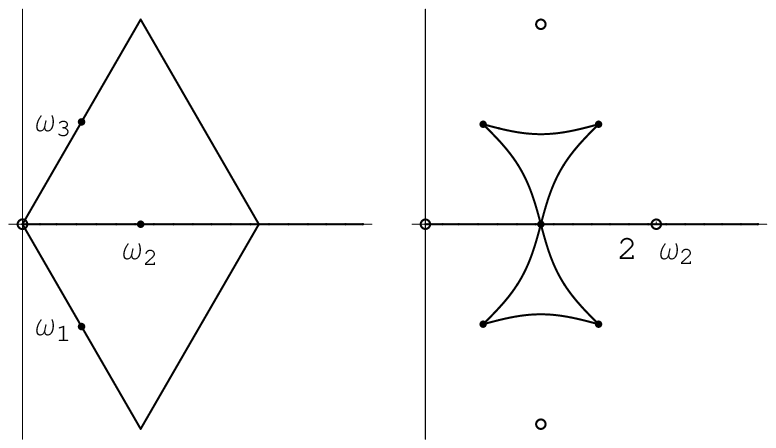,height=3.6cm}}
\vspace{.4cm}
\centerline{\epsfig{file=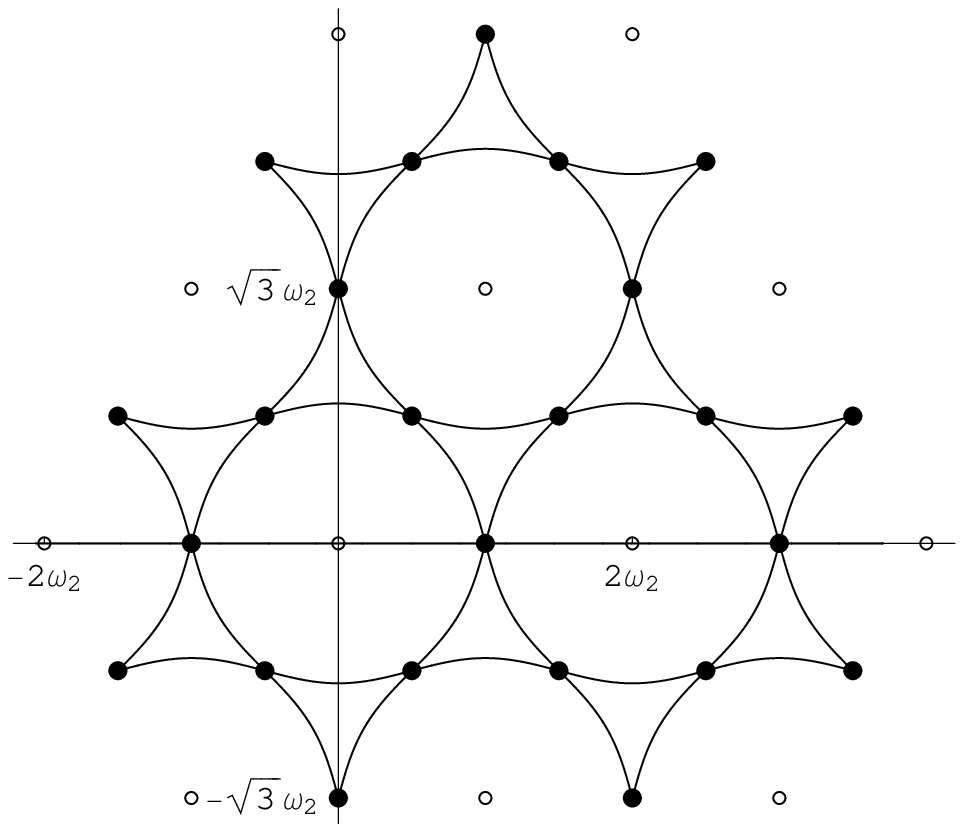,height=3.6cm}}
\caption{The case $N=3,$ showing the minima ${(\bullet)}$ and poles {(o)} of the potential (upper panel, left) and the kink orbits in the FPP (upper panel, right). The lower panel shows the lattice of minima and poles of the deformed potential and the accompanying network of kink orbits.}
\end{figure}

We further illustrate the problem with $N=4.$ Here we have $W(\chi)=\chi-\chi^5/5.$ Thus, the potential is
\be
V(\chi,\ov\chi)=\frac12(1-\chi^4)(1-\ov\chi^4)
\ee
We write $\chi=f(\phi)$ to show that the special deformation function must satisfy
\be\label{eq:dsup1} f'(\phi)
\ov{f'(\phi)}=\,\sqrt{(1-f^4(\phi))(1-\ov{f^4(\phi)})}
\ee
As before, we choose the holomorphic solution of
\be\label{eq:dsup1b}
f^{\prime2}(\phi)=1-f^4(\phi)
\ee
The solution is the elliptic sine of parameter $k^2=-1$, the Gauss's {\it sinus lemniscaticus} $f(\phi)={\rm sn}(\phi,-1).$
The new superpotential is ${\cal W}_\alpha(\phi)=e^{-i \alpha}{\rm sn}(\phi,-1)$ and the deformed potential then reads
\be
U(\phi,\ov{\phi})=\frac12|{\rm cn}(\phi,-1)|^2\cdot |{\rm dn}(\phi,-1)|^2
\ee
The new potential is doubly periodic with an structure inherited from the \lq\lq quarter-periods" $K(-1)={\omega_1/ 4}$ and $i
K(2)={\omega_2/ 4}$ of the twelve Jacobi elliptic functions. Here $K(-1)\approx 1.31103$ is the complete elliptic
integral of the first type, a quarter of the length of the lemniscate curve in field space:
$(\phi_1^2+\phi_2^2)^2=\phi_1^2-\phi_2^2.$ $K(2)\approx 1.31103-i1.31103$ is the complementary complete elliptic integral of $K(-1)$.

The set of zeros of the potential in the FPP are $\phi^{(1)}={\omega_1/4},\phi^{(2)}=i{\omega_1/4},
\phi^{(3)}=-{\omega_1/4},\phi^{(4)}=-i{\omega_1/ 4},$ whereas the set of all the zeros of $U$ form a quadrangular lattice in the whole
configuration space. This is depicted in Fig.~4 and will be fully considered in Ref.~\cite{abglmNPB}. Differently from the former case, however,
here the orbits may connect the minima in two distinct ways: one, with curved lines, in the sequence $(1,2)$, $(2,3)$, $(3,4)$, and $(4,1)$, and the other
with straight line segments, in the sequence $(1,3)$ and $(2,4)$ -- see \cite{abglmNPB} for further details on this issue.

\begin{figure}[htbp]
\centerline{\epsfig{file=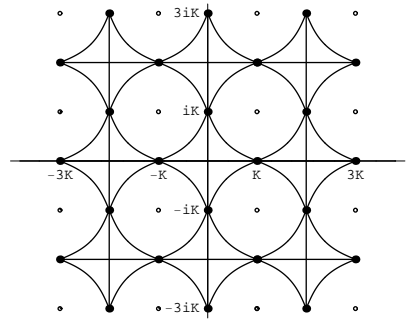,height=4.2cm}}
\caption{The case $N=4,$ showing the set of minima $(\bullet)$ and poles ({o}) of the deformed potential and the accompanying network of kink orbits.}
\end{figure}

In summary, in this work we have used the procedure developed in \cite{DD,EDD} to deform a given model in a way such that its set of minima could be replicated in the entire field space. The idea was developed in the real line, for the case of a real field, and in the plane, for the case of a complex field. Since the set of minima are connected by algebraic orbits describing defect structures in field space, we have also been able to replicate the algebraic orbits in the entire field space of the deformed model, naturally building networks of defects, which are spread from rudimentary networks into the entire field space.

A natural extension of this work concerns the construction of irregular lattices and networks in the plane, in the case of a complex field, which we will study in our next work, now under preparation \cite{abglmNPB}. Another extension concerns the use of three real fields, to investigate if it is possible to tile the space in a way similar to the case of planar networks here considered. 

We recall that a kink-like defect in general splits the space into two distinct regions, so we could also think as in \cite{DW}, using two spatial dimensions, to see how the kinks orbits that we have just obtained could tile the plane with regular and/or irregular polygons, with triple junctions for $N=3$, and with quartic junctions for $N=4$. Another interesting issue could address the same problem, but now embedding the scalar fields in a curved space-time, following the lines of Ref.~{\cite{sugra}}. This would lead us to another route, in which we could try to understand how the networks here introduced would change in a curved background. We can also think of making the space-time dynamically curved, to see how the domain wall networks could follow the evolution investigated in \cite{ammmo}. These and other related issues are presently under consideration, and we hope to report on the them in the near future.

This work is part of a collaboration which has been financed by the Brazilian and Spanish governments: VIA, DB and LL thank CAPES, CLAF, CNPq and PRONEX-CNPq-FAPESQ, and MAGL and JMG thank Ministerio de Educacion y Ciencia, under grant FIS2006-09417, for partial support.


\end{document}